# Revealing 3D Magnetization of Thin Films with Soft X-ray Tomography: Closure Domains and Magnetic Singularities

A. Hierro-Rodriguez[1,*], C. Quirós[2,3], A. Sorrentino[4], L. M. Alvarez-Prado[2,3], J. I. Martín[2,3], J. M. Alameda[2,3], S. McVitie[1], E. Pereiro[4], M. Vélez[2,3,*] and S. Ferrer[4,*]

[1]SUPA, School of Physics and Astronomy, University of Glasgow G12 8QQ, UK.

[2]Depto. Física, Universidad de Oviedo, 33007 Oviedo, Spain.

[3]CINN (CSIC – Universidad de Oviedo), 33940 El Entrego, Spain.

[4]ALBA Synchrotron, 08290 Cerdanyola del Vallès, Spain.

*Correspondence to: Aurelio.HierroRodriguez@glasgow.ac.uk, mvelez@uniovi.es, ferrer@cells.es.

**Abstract:** The knowledge of how the magnetization looks inside a ferromagnet is often hindered by the limitations of the available experimental methods that are sensitive only to the surface regions or limited in spatial resolution. We report the 3D tomographic reconstruction of the magnetization within a ferromagnetic film of 240 nm in thickness using soft X ray microscopy and magnetic dichroism. The film has periodic magnetic domains forming stripes and closure domains found to be shifted from the stripe array by ¼ of the period. In addition, the bifurcations of the stripes, which act as inversion nuclei of the magnetization, evidence in 3D meron singularities and Bloch points at the interior of the film. This novel method can be easily extended to magnetic stacks in spintronics applications and other singularities in films.

Due to the importance of domains in the magnetic properties of materials, including thin films or nanostructures for applications in spintronics, magnetic domain visualization methods have been an active field of research in the last decades [(1,2)]. Transmission methods are excellent for the visualization of magnetic states through the sample and require probes with high enough penetration depths. For instance, neutron radiographies have allowed imaging the interior of ferromagnets at sub mm scale (3,4). Transmission electron microscopy may probe films of thickness up to 100nm (5), which has been used to characterize in-plane magnetization by electron holography (6) and to tomographically reconstruct the magnetic vector potential of individual nanoelements (7). More recently, the magnetic configuration within a 5μm diameter $GdCo_2$ micropillar has been reconstructed by magnetic vector ptycho-tomography using hard X-rays (wavelength 0.17 nm) (8). This has been the first experimental realization of a complete tomographic reconstruction using all the different angular projections to recover the 3D magnetization configuration. In the soft X-ray range, 2D magnetization patterns were extracted from transmission X-ray microscopy images of tubular samples acquired at different angles (9). This approach was based on exploiting the small thickness of the tubular shell to recover, via a single angular series around a particular rotation axis, the magnetic state of the system. Also, operating in reflection mode, detailed characterization of skyrmions in periodic arrays (10-11) has been achieved by soft X ray magnetic scattering.

Tomographic imaging of extended thin samples is of particular difficulty due to the high aspect ratio (lateral dimensions >> thickness). This has hindered a direct 3D visualization of the magnetization in thin films and multilayers of arbitrary magnetization configuration and is a major limitation since most magnetic devices are fabricated on substrates with macroscopic lateral dimensions. Therefore, a variety of basic and application-related topics such as the depth dependence of magnetic textures in chiral multilayers (12), or the optimization of spin-torque oscillators (13) have been addressed indirectly (12-17) by image simulations and clever sample designs. Thus, magnetic vector tomography in extended thin films can become an essential tool for the evolution of nanomagnetism from 2D configurations to the complex 3D magnetization textures and structures explored nowadays (2).

We present here the first results of a 3D tomographic reconstruction of a magnetic thin film heterostructure using transmission soft X-ray microscopy and a recently developed vector reconstruction method (18) allowing us to get insight on the configuration of the 3D magnetization vector **m**. X-ray magnetic circular dichroism (XMCD) provides the contrast mechanism that allows revealing the vector nature of the magnetization: the dichroic effect of a volume element in the material depends on the dot product **σ•m** of the spin angular momentum of the circularly polarized photons **σ** (parallel/antiparallel to the propagation direction for clockwise (CW) or counter-clockwise (CCW) polarizations) and the local magnetization **m** (19). By exploiting the angular dependence of **σ•m** and using appropriate algorithms, the vector reconstruction of the magnetization is obtained. In short, the reconstruction method is based on processing tomograms formed by transmission X-ray micrographs of the sample acquired under different angular orientations. The intensity recorded at each pixel of the detector depends on the line integral of **σ•m** along a specific X-ray path through the sample. By creating a volume model composed of voxels, the line integral can be discretized into a linear equation where the unknowns are proportional to the magnetization vectors within each voxel along the X-ray path. Considering all the pixels and all the different projections, a system of linear equations is constructed. An iterative method based on an Algebraic Reconstruction Technique is used to solve the system of equations (20) obtaining the magnetization vectors. The limitations of the method arise from the extended nature of the film and from the experimental set-up: The effective film thickness increases at grazing angles preventing photon transmission and geometrical shadowing effects limit the measurable angular range (in this case ±55°) causing the so called "missing wedge" (details in (26)).

We fabricated a magnetic trilayer $Ni_{80}Fe_{20}(80nm)/NdCo_5(80nm)/Ni_{80}Fe_{20}(80nm)$ (Permalloy: Py = $Ni_{80}Fe_{20}$) by DC magnetron sputtering at room temperature on top of a 50nm thick $Si_3N_4$ membrane to test the capabilities of the method. The $NdCo_5$ layer (21) displays weak perpendicular magnetic anisotropy (PMA) leading to magnetic stripe domains with canted up and down magnetizations (22). The exchange interaction imprints the central stripe pattern to the magnetically soft Py films. Top and bottom layers are made of the same material so that they are indistinguishable by standard X-ray transmission microscopy. Although micrographs with good magnetic contrast have been achieved in Py films only 10 nm thick, we report here results on thicker samples to probe the in–depth magnetization. The depth resolution is approximately 85 nm (details in (26)) and the maximum film thickness compatible with reasonable transmission is 300-400nm.

The sample was prepared for the experiment by initiating the in-plane magnetization reversal after saturation to promote the formation of non-trivial magnetic textures (23,24), and then it was loaded in the full-field X-ray transmission microscope of the Mistral beamline at the Alba synchrotron (25) [Fig. 1(a)]. The sample, mounted in a high precision rotary stage, was illuminated with circularly polarized X-rays with handedness (CW or CCW) selected by a set of movable ancillary slits. The XMCD effect probes the magnetization in a plane perpendicular to the rotation axis ($\theta$ rotation). Hence, in the sample's reference frame [Fig. 1(b)], the in-plane magnetization perpendicular to the rotation axis ($m_{//}$) and the out-of-plane component ($m_\perp$) can be measured. The magnetic contrast is separated from the charge contribution by subtracting CW and CCW images under the same sample angular orientation. At the Fe $L_3$ absorption energy edge (wavelength 1.754 nm), the images probe the magnetization of the Py layers whilst no magnetic signal is originated from the central NdCo layer. Figs.1(c-d) show micrographs acquired for two different azimuthal orientations (Tilt series 1 and 2) allowing us to probe the three components of **m**. Tilt series 1 [Fig. 1(c)] is mostly sensitive to magnetization along the stripes ($m_{//}1$) and normal to the film ($m_\perp 1$). Tilt

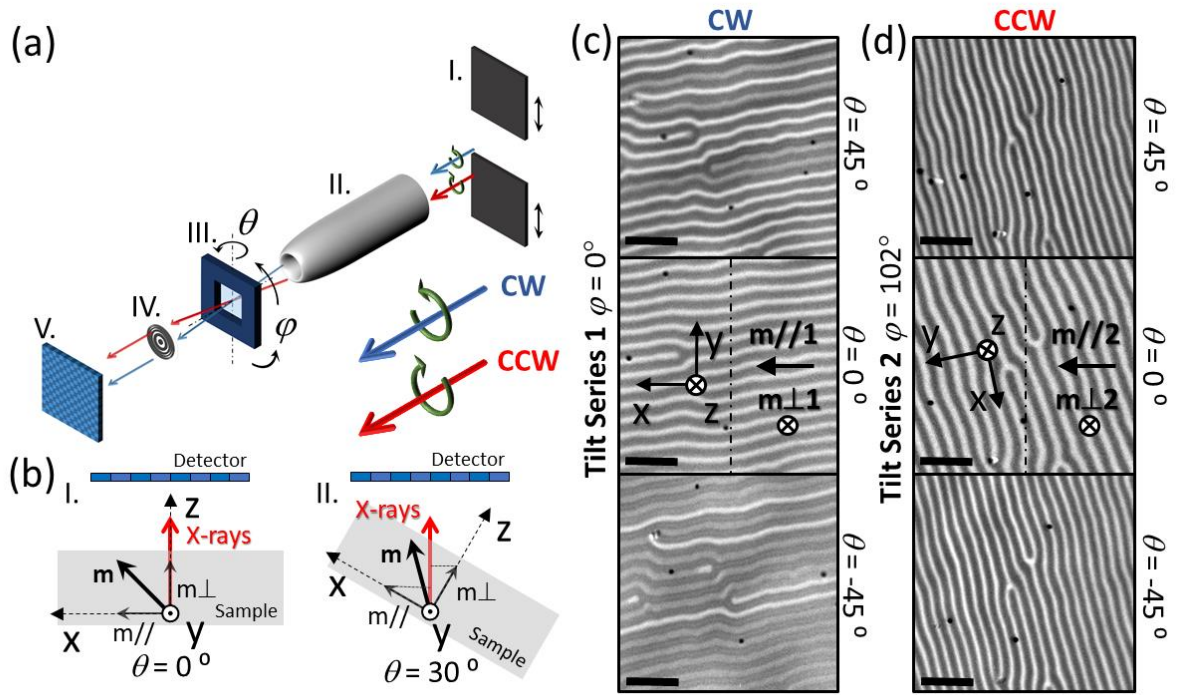

**Figure 1** (a) Scheme of the full field soft X-ray transmission microscope: (I) Ancillary slits to select the polarization of the X rays (circular CW or CCW), (II) capillary condenser, (III) goniometric stage ($\theta$ rotation), $\varphi$ rotation manually performed outside the microscope, (IV) Fresnel zone plate and (V) CCD detector. (b) Top view sketch to illustrate sample reference frame and magnetization components probed during a $\theta$ rotation (around the y axis): (I) Normal incidence ($\theta=0°$) leads to out-of-plane ($m\perp$) sensitivity. (II) Oblique incidence ($\theta=30°$) allows for in-plane ($m//$) and out-of-plane ($m\perp$) sensitivity. (c-d) Experimental images for two different sample configurations,: (c) Tilt series 1 ($\varphi=0°$) for CW polarization, (d) Tilt series 2 ($\varphi=102°$) for CCW polarization. Reference frame indicated in both datasets. ($m//$, $m\perp$) indicate probed magnetization components. Dot-dash line indicates $\theta$ rotation axis. Scale bars 1μm.

series 2 [Fig. 1(d)] probes mainly the magnetization transversal to the stripes ($m//2$) and normal to the surface ($m\perp 2$). Tilt series 2 was acquired after a manual rotation of the sample of 102° [φ rotation in Fig. 1(a)]. The black dots scattered across the images are 100nm gold nanoparticles used as fiducial markers for accurate projection alignment prior to reconstruction. These are essential for tomography of thin films since the sample shape is effectively an infinite plane and it cannot be used for alignment as done previously (8,9). Each Tilt series is composed by a set of 87 projections collected by varying the angle of incidence (θ) for both photon polarizations.

Note the inverted magnetic contrast for both polarizations at θ=0° [Fig. 1(c-d)], and the partial inversion occurring between θ=±45° in Tilt series 1, which result from the different signs of **σ·m**. See supplementary material (26) for further experimental details.

In the following we will discuss first the tomographic reconstruction of each separated Tilt series and finally merge the information of both for the final reconstruction. The first approach aims to analyze the effects of the thin film geometry on the resulting magnetization, while the latter recovers the full 3D magnetization structure of the Py layers.

To analyze the reconstructed volume (400x400x80 voxels, cell-size 10.5nm), the central slices of m// and m⊥ magnetization components of top and bottom Py layers [Fig. 2(a)] for both Tilt series have been extracted. The m//1 images [Fig. 2(b)] present an oscillatory dark contrast flecked with brighter stripes indicating the beginning of the in-plane magnetization reversal that is different at top and bottom slices in the vicinity of the bifurcations (D1 and D2). m⊥1 and m⊥2 images are very similar, as both are sensitive to the same magnetization component, and display identical alternating contrasts in the top and bottom slices. Finally, the contrast of m//2 exhibits periodic bright-dark oscillations that indicate a periodic change of sign in the magnetic component transversal to the stripes. A deeper insight is obtained by representing the cross sections of the reconstructed datasets indicated by the red arrows in the top-view images. Here, a major difference is observed between the two Tilt series: the cross sections for m//1 and m⊥1 show magnetic contrast extended through the whole thickness of the reconstruction volume (850nm) [Fig. 2(d)] while m//2 and m⊥2 allocate the magnetization in a confined region in z [Fig. 2(e)]. The reason for this difference is a parallax effect due to the different orientations of the stripes with the rotation axis. As visible in Fig. 1 at φ=102°, the

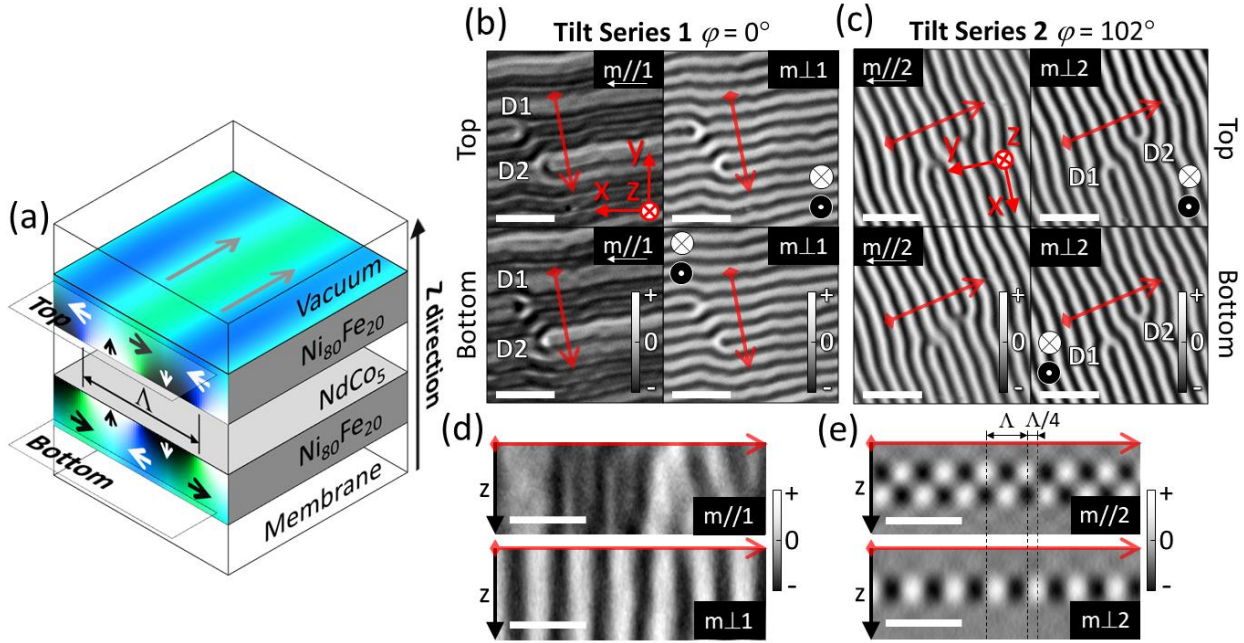

**Figure 2** (a) Sketch of stripe domain magnetization in NiFe/NdCo/NiFe trilayer. (b-c) Reconstructed magnetization slices at top and bottom Py layers showing in-plane (m//) and out-of-plane (m⊥) components for Tilt series 1 (b,φ=0°) and 2 (c,φ=102°). D1 and D2 index the bifurcations in the image. Red arrows indicate the extracted cross sections. Greyscale bars indicate the magnetization sign. (d-e) Cross sections of in-plane (m//) and out-of-plane (m⊥) components of the reconstructed magnetization for Tilt series 1 (d) and 2 (e). Vertical dashed lines in (e) indicate the stripe periodicity (Λ) and the Λ/4 dephasing in between m//2 and m⊥2 components in the closure domain structure. Scale bars (b-c) 1.4μm and (d-e) 700nm.

stripes appear thinner and closer at oblique angles whereas at φ=0° this does not occur since the stripes are much longer than the acceptance of the detector originating the spread along z in Fig. 2(d). The reconstructed signal from Tilt series 2 locates the magnetization in a z-extension of ~300nm, which agrees with the nominal thickness of the film (240nm) considering the estimated z resolution ~85nm. It is important to mention here that our tomographic reconstruction does not show zero magnetization at the central NdCo layer (invisible at the Fe L3 photon energy) but instead, it displays a continuous evolution from the top Py film to the bottom one. Most probably this is due to the missing wedge effect originated by the limited angular range accessible. Nevertheless, this can be exploited to get insight on the magnetic configuration within the central layer since the exchange interaction ensures a continuous evolution of the magnetization between both Py layers.

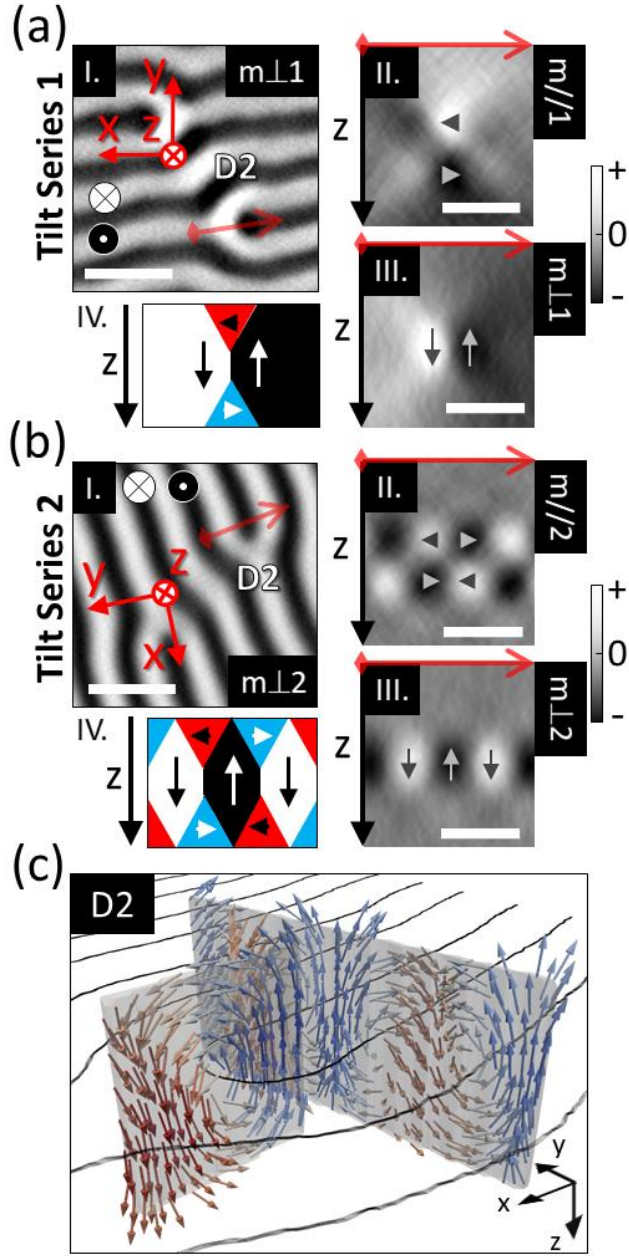

**Figure 3** (a-b) Reconstructed magnetization around bifurcation D2 for Tilt series 1 (a) and 2 (b). (I.) Top view of m⊥ at central slice of volume model. Greyscale bars indicate the sign of the magnetization. Red arrows indicate the extracted cross-sections. Cross-sections show in-plane (II., m//) and out-of-plane (III., m⊥) components for Tilt series 1 (a) and 2 (b). (IV.) Sketches in (a-b) show the closure magnetization structure (c) 3D vector representation of the reconstructed magnetization from the selected cross-sections. Black lines indicate the Bloch domain-walls. Scale bars (a-b) 700nm (top-view) and 350nm (cross-section).

Note that in Fig. 2(e), positive m//2 regions of the upper Py layer are on top of negative regions of the bottom layer. This is due to the circulating magnetization of the closure domains or Neel caps (22) characteristic of stripe domains, sketched in Fig. 2(a). Note the Λ/4 dephasing (Λ≈390nm determined from the data) between the in-plane and out of plane magnetization oscillation as indicated by the dashed vertical lines, typical of the closure domain structure.

We focus now on the magnetization around the bifurcations within the stripe pattern as they are key actors in the in-plane magnetization inversion (23,24,28). Specifically, we analyze the reconstructed magnetization around the D2 bifurcation by extracting longitudinal [Fig. 3(a)I., red arrow] and transversal [Fig. 3(b)I., red arrow] cross sections. The first observation is that we obtain good confinement of the magnetization at the bifurcation core for m//1 and m⊥1 [Fig. 3(a)II.-III.] due to the smaller parallax at this position than along the stripes. Moreover, the magnetization senses resulting from the signs for both Tilt series display the characteristic circulation vortices from top to bottom around the defect. From left to right for Tilt series 1, a m⊥1 positive stripe evolves into a negative one through the core of the dislocation showing positive (negative) m//1 component at bottom (top) Py layers [Fig. 3(a)II.-IV.]. Also, from left to right, a m⊥2 positive component stripe changes into a negative one and back to positive along the displayed cross section, by negative (positive) and positive (negative) m//2 closure Neel caps within the top (bottom) magnetic layers [Fig. 3(b)II.-IV.]. This is further clarified in Fig. 3(c) where both cross sections are shown in vector representation with the Bloch domain-wall (black lines) separating different oriented stripes.

Finally, we performed the joint reconstruction using the data from both Tilt series to obtain the three-dimensional magnetization at the core of dislocations. It was achieved by a careful alignment of the projections of both Tilt series using the gold fiducials (visible in Fig. 1(c-d) as black points). This allowed us to reconstruct a single magnetization model of the three components of the magnetization (details in (26)). Figure 4(a) shows the $m_z$ component at the central slice of the reconstructed volume that has dislocations D1 and D2.

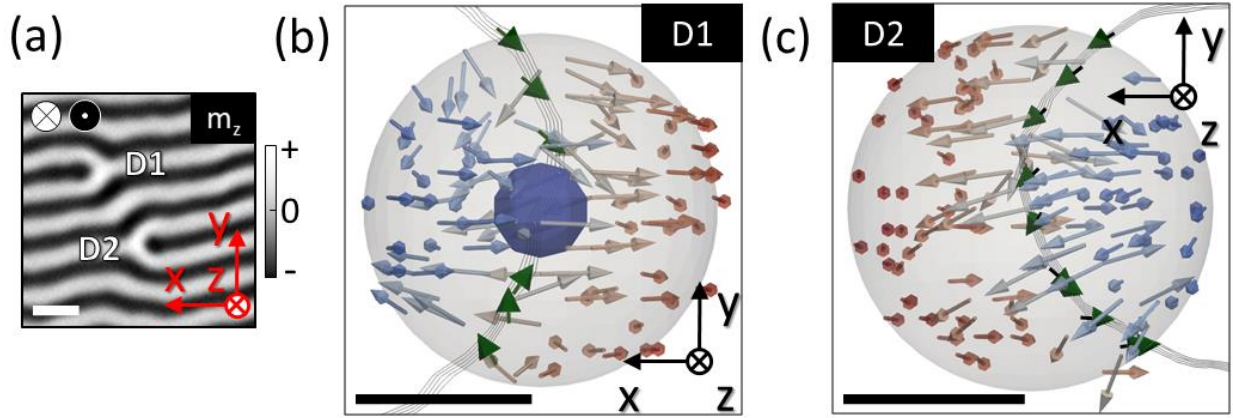

**Figure 4** (a) Out-of-plane ($m_z$) component from the joint reconstruction of Tilt-series 1 and 2 at the central slice of the volume model. Dislocations D1 and D2 are indicated. Grey scale bar indicates the sign of $m_z$. (b-c) 3D magnetic textures at the cores of D1 (b) and D2 (c). Arrows depict the orientation of the magnetization colored with Blue-White-Red indicating Negative-Zero-Positive sign of $m_z$. Green arrows show the magnetization within the Bloch domain-wall at the core of the dislocations. Blue sphere marks the Bloch-point singularity. Scale bar (a) 525nm, (b) 75nm, (c) 105nm.

Both defects look very similar in Fig. 4(a), however, while bifurcation D1 displays a convergent in-plane magnetization towards the core of the defect in Fig. 4(b) [$m_x$ keeps its negative sign and goes to zero while $m_y$ has positive (negative) value at the lower (upper) branch of the Bloch domain-wall in the proximity of the dislocation core], bifurcation D2 exhibits a continuous CCW circulation following the domain-wall [Fig. 4(c)]. This is consistent with a meron-like configuration with topological charge ½ (27,28), while Fig. 4(b) strongly suggests the existence of a circulating Bloch-point at the core of the dislocation (8,29). The actual position of the Bloch point is depicted by the blue sphere in Fig. 4(b) since it cannot be resolved due to its small size and its magnetization-divergent nature. The presence of Bloch-points at bifurcation cores is in agreement with their essential role as drivers of the in-plane magnetization reversal in stripe domain systems (23,24).

In conclusion, we have shown that soft X-ray vector magnetic tomography provides a novel, relatively simple and well-suited method for determining the 3D magnetic configuration of thin films up to 200-300 nm in thickness. The method has been used to map the closure domains in Py films separated by a weak PMA ferromagnetic spacer. Moreover, the method provides insight on the magnetic textures present at the cores of the dislocations identifying a meron-like configuration and inferring a texture compatible with a circulating Bloch point. Further development of the technique exploiting the element sensitivity of the magnetic dichroism will allow resolving the complete configuration of the magnetization in stacks having different magnetic elements at unprecedented detail providing an useful tool in a variety of heterostructures for applications.

**Acknowledgments**: Thanks to R. Valcarcel for technical assistance. A.H.-R. acknowledges the fruitful discussions with G.W. Paterson and thanks C. Donnelly for the useful recommendations about 3D magnetization representation. **Funding**: Alba light source is funded by the Ministry of Research and Innovation of Spain and by the Generalitat de Catalunya. This project has been supported by Spanish MINECO under grant FIS2016-76058 (AEI/FEDER, EU)). A.H.-R. and S.MV. acknowledge the support from European Union's Horizon 2020 research and innovation program under Marie Skłodowska-Curie grant ref. H2020-MSCA-IF-2016-746958.

**References**

1. For a recent review see P. Fischer, X-Ray Imaging of Magnetic Structures. *IEEE Trans. Magn.* **51**, 0800131 (2015).

2. For a recent review see A. Fernández-Pacheco *et al.*, Three-dimensional nanomagnetism. *Nat. Comm.* **8**, 15756 (2017).

3. N. Kardjilov *et al.*, Three-dimensional imaging of magnetic fields with polarised neutrons. *Nat. Phys.* **4**, 399-403 (2008).

4. I. Manke *et al.*, Three-dimensional imaging of magnetic domains. *Nat. Comm.* **1**, 125 (2010).

5. S. McVitie *et al.*, A transmission electron microscope study of Néel skyrmion magnetic textures in multilayer thin film systems with large interfacial chiral interaction. *Sci. Rep.* **8**, 5703 (2018).

6. F. Zheng *et al.*, Experimental observation of chiral magnetic bobbers in B20-type FeGe. *Nat. Nan.* 13, 451 (2018).

7. C. Phatak, A.K. Petford-Long, and M. De Graef, Three-Dimensional Study of the Vector Potential of Magnetic Structures. *Phys. Rev. Lett.* **104**, 253901 (2010).

8. C. Donnely *et al.*, Three-dimensional magnetization structures revealed with X-ray vector nanotomography. *Nature* **547**, 328-331 (2017).

9. R. Streubel *et al.*, Retrieving spin textures on curved magnetic thin films with full-field soft X-ray microscopies. *Nat. Comm.* **6**, 7612 (2015).

10. J. Y. Chauleau *et al.*, Chirality in Magnetic Multilayers Probed by the Symmetry and the Amplitude of Dichroism in X-Ray Resonant Magnetic Scattering. *Phys. Rev. Lett.* **120**, 037202 (2018).

11. S. Zhang *et al.*, Reciprocal space tomography of 3D skyrmion lattice order in a chiral magnet. *Proc. Natl. Acad. Sci. U.S.A.* **115**, 6386-6391 (2018).

12. W. Legrand *et al.*, Hybrid chiral domain walls and skyrmions in magnetic multilayers. *Sci. Adv.* **4**, eaat0415 (2018).

13. S. Klingler *et al.*, Spin-Torque Excitation of Perpendicular Standing Spin Waves in Coupled YIG/Co Heterostructures. *Phys. Rev. Lett.* **120**, 127201 (2018).

14. S. Schneider *et al.*, Induction Mapping of the 3D-Modulated Spin Texture of Skyrmions in Thin Helimagnets. *Phys. Rev. Lett.* **120**, 217201 (2018).

15. D. Song *et al.*, Quantification of Magnetic Surface and Edge States in an FeGe Nanostripe by Off-Axis Electron Holography. *Phys. Rev. Lett.* **120**, 167204 (2018).

16. T. Liu *et al.*, Nontrivial Nature and Penetration Depth of Topological Surface States in $SmB_6$ Thin Films. *Phys. Rev. Lett.* **120**, 207206 (2018).

17. C. Quiros *et al.*, Cycloidal Domains in the Magnetization Reversal Process of $Ni_{80}Fe_{20}/Nd_{16}Co_{84}/Gd_{12}Co_{88}$ Trilayers. *Phys. Rev. Appl*. **10**, 014008 (2018).

18. A. Hierro-Rodriguez *et al.*, 3D reconstruction of magnetization from dichroic soft X-ray transmission tomography. *J. Synchrotron Rad.* **25**, 1144-1152 (2018).

19. J. Stohr, H. C. Siegmann, *Magnetism*, (Springer-Verlag, Berlin-Heidelberg, 2006).

20. A.C. Kak & M. Slaney, *Principles of Computerized Tomographic Imaging*, (New York: IEEE Press, 1988).

21. A. Hierro-Rodriguez, *et al.*, Fabrication and magnetic properties of nanostructured amorphous Nd–Co films with lateral modulation of magnetic stripe period, *J. Phys. D. Appl. Phys* **46**, 345001 (2013).

22. A. Hubert, R. Schäfer, *Magnetic Domains: The Analysis of Magnetic Nanostructures*, (Springer-Verlag, Berlin-Heidelberg, 2006).

23. A. Hierro-Rodriguez, *et al.*, Observation of asymmetric distributions of magnetic singularities across magnetic multilayers. *Phys. Rev. B* **95**, 014430 (2017).

24. A. Hierro-Rodriguez, *et al.*, Deterministic propagation of vortex-antivortex pairs in magnetic trilayers, *Appl. Phys. Lett.* **110**, 262402 (2017).

25. E. Pereiro *et al.*, A soft X-ray beamline for transmission X-ray microscopy at ALBA. *J. Synchrotron Rad*. **16**, 505-512 (2009).

26. Supplementary Material

27. M. Ezawa, Compact merons and skyrmions in thin chiral magnetic films. *Phys. Rev. B* **83**, 100408(R) (2011).

28. C. Blanco-Roldán *et al.*, Nanoscale imaging of buried topological defects with quantitative X-ray magnetic microscopy. *Nat. Comm.* **6**, 8196 (2015).
29. MY. Im, *et al.*, Dynamics of the Bloch point in an asymmetric permalloy disk. *Nat. Comm.* **10**, 593 (2019).